\documentclass[aps,prb,reprint,superscriptaddress]{revtex4-2}
\usepackage{graphicx}
\usepackage{amsmath}
\usepackage{amssymb}
\usepackage{bm}
\usepackage{physics}
\usepackage{subfigure}
\usepackage[colorlinks=true,linkcolor=blue,urlcolor=blue,citecolor=blue]{hyperref}
\usepackage{times}
\usepackage{changes}
\begin{document}

\newcommand{\Hop}{\hat{H}}
\newcommand{\Himp}{\hat{H}_{\rm imp}}
\newcommand{\Uimp}{\hat{U}_{\rm imp}}
\newcommand{\gHimp}{\mathcal{H}_{\rm imp}}
\newcommand{\gUimp}{\mathcal{U}_{\rm imp}}
\newcommand{\Hbath}{\hat{H}_{\rm bath}}
\newcommand{\Hhyb}{\hat{H}_{\rm hyb}}

\newcommand{\aop}{\hat{a}}
\newcommand{\adop}{\hat{a}^{\dagger}}
\newcommand{\sgp}{\hat{\sigma}^+}
\newcommand{\sgx}{\hat{\sigma}^x}
\newcommand{\sgy}{\hat{\sigma}^y}
\newcommand{\sgz}{\hat{\sigma}^z}
\newcommand{\nop}{\hat{n}}

\newcommand{\cop}{\hat{c}}
\newcommand{\cdop}{\hat{c}^{\dagger}}
\newcommand{\hc}{{\rm H.c.}}
\newcommand{\rhotot}{\hat{\rho}_{\mathrm{tot}}}
\newcommand{\rhoop}{\hat{\rho}}
\newcommand{\rhoimp}{\hat{\rho}_{\mathrm{imp}}}
\newcommand{\rhobath}{\hat{\rho}_{\mathrm{bath}}}
\newcommand{\Zimp}{Z_{{\rm imp}}}
\newcommand{\mea}{\mathcal{D}}
\newcommand{\gK}{\mathcal{K}}
\newcommand{\gI}{\mathcal{I}}
\newcommand{\gF}{\mathcal{F}}
\newcommand{\bolda}{\bm{a}}
\newcommand{\boldabar}{\bar{\bm{a}}}
\newcommand{\abar}{\bar{a}}
\newcommand{\im}{{\rm i}}
\newcommand{\contour}{\mathcal{C}}
\newcommand{\gA}{\mathcal{A}}
\newcommand{\gB}{\mathcal{B}}
\newcommand{\gM}{\mathcal{M}}
\newcommand{\boldeta}{\bm{\eta}}
\newcommand{\boldetabar}{\bar{\bm{\eta}}}
\newcommand{\etabar}{\bar{\eta}}
\newcommand{\parity}{\mathcal{P}}
\newcommand{\current}{\mathcal{J}}
\newcommand{\pronyerror}{\varsigma_p}

\newcommand{\WI}{{\rm W}^I}
\newcommand{\WII}{{\rm W}^{II}}

\newcommand{\EqDef}{\stackrel{\mathrm{def}}{=}}

\newcommand{\gcc}[1]{{\color{red}#1}}
\definechangesauthor[color=orange]{RF}

\title{Infinite Grassmann time-evolving matrix product operator method for zero-temperature equilibrium quantum impurity problems}


\author{Chu Guo}
\email{guochu604b@gmail.com}
\affiliation{Key Laboratory of Low-Dimensional Quantum Structures and Quantum Control of Ministry of Education, Department of Physics and Synergetic Innovation Center for Quantum Effects and Applications, Hunan Normal University, Changsha 410081, China}

\author{Ruofan Chen}
\email{physcrf@sicnu.edu.cn}
\affiliation{College of Physics and Electronic Engineering, and Center for Computational Sciences, Sichuan Normal University, Chengdu 610068, China}

\date{\today}

\begin{abstract}
The Grassmann time-evolving matrix product operator (GTEMPO) method has proven to be an accurate and efficient numerical method for the real-time dynamics of quantum impurity problems. 
Whereas its application for imaginary-time calculations is much less competitive compared to well-established methods such as the continuous-time quantum Monte Carlo (CTQMC). 
In this work, we unleash the full power of GTEMPO for zero-temperature imaginary-time calculations: the multi-time impurity state is time-translationally invariant with infinite boundary condition, therefore it can be represented as an infinite Grassmann matrix product state (GMPS) with nontrivial unit cell in a single time step, instead of an open boundary GMPS spanning the whole imaginary-time axis. We devise a very efficient infinite GTEMPO algorithm targeted at zero-temperature equilibrium quantum impurity problems, which is known to be a hard regime for quantum Monte Carlo methods. 
To demonstrate the performance of our method, we benchmark it against exact solutions in the noninteracting limit, and against CTQMC calculations in the Anderson impurity models with up to two orbitals, where we show that the required bond dimension of the infinite GMPS is much smaller than its finite-temperature counterpart.
\end{abstract}
\maketitle

\section{Introduction}

The quantum impurity problem (QIP) considers a small impurity consisting of a few energy levels which is coupled to a continuous bath of free particles. 
QIP plays a fundamental role in quantum physics for studying non-Markovian open quantum effects~\cite{LeggettZwerger1987} and strongly correlated effects~\cite{mahan2000-many}. It is also a building block for quantum embedding methods such as the dynamical mean field theory~\cite{GeorgesRozenberg1996}. 
Mathematically, the QIPs are described by a very special type of sparse Hamiltonian in which the integrability is only broken by the presence of interaction inside the impurity. Despite this formal simplicity, accurate numerical solutions to QIPs are generally challenging and 
there is a continuing effort to develop efficient impurity solvers due to their profound importance.

The focus of this work is on the numerically exact solutions to the fermionic QIPs, primarily the Anderson impurity models (AIMs)~\cite{anderson1961-localized}, but the method developed in this work can also be straightforwardly generalized to the bosonic QIPs.
For AIMs, the class of observables of vital interest is the spectral function, which is the imaginary part of the retarded Green's function defined in the real-frequency axis. 
In practice, the calculation of the spectral function is often mitigated to calculating the Matsubara Green's function defined in the imaginary-frequency axis, since the latter can generally be calculated more efficiently,
and the spectral function can be obtained from the Matsubara Green's function via analytical continuation in principle. A number of computational techniques have been developed for the imaginary-time calculations, such as the continuous-time quantum Monte Carlo (CTQMC) method which is based on the perturbative expansion of the Matsubara Green's function~\cite{GullWerner2011,RubtsovLichtenstein2005,GullTroyer2008,WernerMillis2006b,WernerMillis2006,ShinaokaWerner2017,EidelsteinCohen2020}, and the wave-function based methods which explicitly discretize the bath and then perform imaginary-time evolution for the impurity-bath wave function, including exact diagonalization~\cite{CaffarelKrauth1994,KochGunnarsson2008,GranathStrand2012,LuHaverkort2014,ZaeraLin2020,LuHaverkort2019,HeLu2014,HeLu2015}, the numeric renormalization group~\cite{Wilson1975,Bulla1999,BullaPruschke2008,Frithjof2008,ZitkoPruschke2009,DengGeorges2013,StadlerWeichselbaum2015,LeeWeichselbaum2016,LeeWeichselbaum2017}, and matrix product state (MPS) based methods~\cite{WolfSchollwock2014b,GanahlEvertz2014,GanahlVerstraete2015,WolfSchollwock2015,GarciaRozenberg2004,NishimotoJeckelmann2006,WeichselbaumDelft2009,BauernfeindEvertz2017,WernerArrigoni2023,KohnSantoro2021,KohnSantoro2022}. For general equilibrium AIMs, currently the CTQMC methods are the method of choice due to their well-balanced accuracy and computational efficiency, despite the sign problem~\cite{TroyerWiese2005} and the unavoidable sampling noises. In comparison, for practical applications the wave-function based methods could often suffer from inaccuracy or scalability issues resulting from the bath discretization.

The time-evolving matrix product operator (TEMPO) method is a recently emerged non-wave-function based approach, which makes use of the analytical solution of the Feynman-Vernon influence functional (IF)~\cite{FeynmanVernon1963} to integrate out the bath exactly, and then represents the multi-time impurity state as a matrix product state (MPS)~\cite{StrathearnLovett2018}. TEMPO is originally developed for bosonic QIPs~\cite{StrathearnLovett2018,JorgensenPollock2019,popovic2021-quantum,fux2021-efficient,gribben2021-using,otterpohl2022-hidden,gribben2022-exact,chen2023-heat} and recently extended to AIMs by us under the name of Grassmann TEMPO (GTEMPO), since Grassmann MPS (GMPS) is used to deal with the Grassmann path integral (PI) for fermionic QIPs~\cite{ChenGuo2024a,ChenGuo2024b,ChenGuo2024c}. For AIMs a tensor network IF method is also developed recently which makes use of the MPS representation of the Feynman-Vernon IF in the Fock state basis~\cite{ThoennissAbanin2023a,ThoennissAbanin2023b,NgReichman2023}. To date, the (G)TEMPO methods have established themselves as one of the most competitive methods for studying the real-time dynamics of QIPs, most prominently the non-equilibrium quantum transport~\cite{chen2023-heat,ChenGuo2024a}. 
However, when applied for equilibrium AIMs, it has been observed for both the GTEMPO~\cite{ChenGuo2024b} and the tensor network IF~\cite{KlossAbanin2023} methods that the required number of states needs to be kept (e.g., the bond dimension) in the MPS grows much faster (close to linear scaling against inverse temperature $\beta$) than in the real-time evolution. This observation contradicts with almost all the previous methods, where it has been well-established that imaginary-time calculations are generally more efficient than real-time calculations. The origin of this counter-intuitive result is two-fold~\cite{ChenGuo2024b}: (1) the hybridization function in the exponent of the IF contains a growing contribution for finite-temperature equilibrium AIMs; (2) the imaginary-time PI is time-translationally invariant (TTI) with anti-periodic boundary condition, but is represented as an open boundary MPS in existing GTEMPO or tensor network IF studies.

In this work we unleash the full power of the GTEMPO method for zero-temperature equilibrium AIMs by exploring its TTI property: the underlying multi-time impurity state can be naturally represented as an infinite GMPS where only the site tensors in a single time step are independent. 
Moreover, in this case the hybridization function only contains decaying contributions, which promises an efficient GMPS representation of the IF (referred to as MPS-IF afterwards) with much lower bond dimension compared to the finite-temperature case. 
Correspondingly, we will refer to this approach as the infinite GTEMPO (iGTEMPO) method throughout this work.
We numerically confirm the small bond dimension of the zero-temperature MPS-IF for the commonly used semi-circular bath spectrum density, and we also demonstrate the accuracy and flexibility of the iGTEMPO method against the low-temperature CTQMC calculations for AIMs with up to two orbitals.
Our results show that
the iGTEMPO method could be highly competitive for zero temperature imaginary-time calculations, which is a hard regime for CTQMC. 

\section{The imaginary-time path integral formalism}\label{sec:PI}
The Hamiltonian of the multi-orbital Anderson impurity model can be generally written as $\Hop = \Himp + \Hbath + \Hhyb$,
with $\Himp=\sum_{p, q} t_{p, q} \adop_{p}\aop_{q} + \sum_{p,q,r,s} v_{p,q,r,s} \adop_p\adop_q\aop_r\aop_s$ the impurity Hamiltonian, $\Hbath=\sum_{p, k} \varepsilon_{p, k} \cdop_{p, k} \cop_{p, k}$ the bath Hamiltonian and $\Hhyb=\sum_{p, k} V_{p, k}(\adop_{p} \cop_{p, k} + \cdop_{p, k}\aop_{p} )$ the hybridization Hamiltonian coupling the impurity to the bath. In our notation $\aop$ and $\adop$ are the fermionic annihilation and creation operators, $p,q,r,s$ are fermion flavor indices of the impurity which contain both the spin and orbital indices, $k$ is the momentum index of the itinerant fermions, $\varepsilon_{p, k}$ denotes the band energy and $V_{p, k}$ is the hybridization strength. The noninteracting bath and the form of linear coupling in $\Hhyb$ ensure that the analytical expression of the Feynman Vernon IF (thus the GTEMPO method) can be applied.
For notational convenience we assume $\Hbath$ and $\Hhyb$ to be flavor-independent in the following, namely $\varepsilon_{p,k} = \varepsilon_k$ and $V_{p, k} = V_k$.

For equilibrium AIMs with inverse temperature $\beta$, the impurity partition function $\Zimp \EqDef \Tr e^{-\beta\Hop} / \Tr e^{-\beta\Hbath}$ can be written as a path integral~\cite{kamenev2009-keldysh,negele1998-quantum}:
\begin{align}\label{eq:PI}
  \Zimp=\int\mea[\boldabar,\bolda]\gK[\boldabar,\bolda]\sum_p \gI_p[\boldabar_p,\bolda_p],
\end{align}
where $\boldabar_p=\{\abar_p(\tau)\}$ and $\bolda_p=\{a_p(\tau)\}$ are Grassmann trajectories for flavor $p$ over the continuous imaginary-time interval $[0, \beta]$, $\boldabar = \{\boldabar_p, \boldabar_q, \cdots\}$ and $\bolda = \{\bolda_p, \bolda_q, \cdots\}$ are abbreviations for Grassmann trajectories of all flavors. 
The measure is defined as $\mea[\boldabar,\bolda]=\prod_{p, \tau}\dd\abar_p(\tau)\dd a_p(\tau)e^{-\abar_p(\tau)a_p(\tau)}$.
The bare impurity dynamics is encoded in $\gK$, determined only by $\Himp$.
The influence of the bath on the impurity is encoded in the IF $\gI_p$ for each flavor, which can be written as 
\begin{align}
  \label{eq:IF}
  \gI_p[\boldabar_p,\bolda_p]=e^{-\int_0^{\beta}\dd\tau\int_0^{\beta}\dd\tau'\abar_p(\tau)\Delta(\tau,\tau')a_p(\tau')}.
\end{align}
The hybridization function $\Delta(\tau,\tau')$ in the double integral of Eq.(\ref{eq:IF}) full characterizes the bath effects, which is defined as
\begin{align}\label{eq:hybrid}
  \Delta(\tau,\tau')=\int\dd{\varepsilon}J(\varepsilon)D_{\varepsilon}(\tau,\tau'),
\end{align}
with $J(\varepsilon)$ the bath spectrum density. $D_{\varepsilon}(\tau,\tau')$ is the free bath Matsubara Green's function, defined as
\begin{align}\label{eq:freegf}
  D_{\varepsilon}(\tau,\tau')\EqDef &-\expval*{T_{\tau}\cop_{\varepsilon}(\tau)\cdop_{\varepsilon}(\tau')}_0 \nonumber \\ 
  =&-[\Theta(\tau-\tau')-n_F(\varepsilon)]e^{-\varepsilon(\tau-\tau')},
\end{align}
where $\Theta$ is the Heaviside step function, and $n_F(\varepsilon)=(e^{\beta\varepsilon}+1)^{-1}$ is the Fermi-Dirac distribution.
For finite $\beta$, the first term on the second line of Eq.(\ref{eq:freegf}) is nonzero for any value of $\varepsilon$, while the exponent of the second term, $-\varepsilon(\tau-\tau')$, can be both positive and negative, therefore $D_{\varepsilon}(\tau,\tau')$ contains both exponentially decaying and growing contributions. 
However, in the zero-temperature limit, $D_{\varepsilon}(\tau,\tau')$ can be written as 
\begin{align}\label{eq:freegf0}
D_{\varepsilon}(\tau,\tau')\vert_{\beta=\infty} = \begin{cases}
  -e^{-\varepsilon(\tau-\tau')},  &\text{if } \varepsilon > 0, \tau \geq \tau'; \\
  0, &\text{if } \varepsilon > 0, \tau < \tau'; \\
  0,  &\text{if } \varepsilon < 0, \tau \geq \tau'; \\
  e^{-\varepsilon(\tau-\tau')},  &\text{if } \varepsilon < 0, \tau < \tau', \\
\end{cases}
\end{align}
where the only two nonvanishing terms on the rhs are purely decaying, as their exponents are both negative.


\section{Method description}

\subsection{The time-translational invariance of the imaginary-time PI}

The integrand of the PI in Eq.(\ref{eq:PI}), denoted as 
\begin{align}\label{eq:adt}
\gA[\boldabar, \bolda] \EqDef \gK[\boldabar,\bolda]\sum_p \gI_p[\boldabar_p,\bolda_p],
\end{align}
represents the multi-time impurity state, which is referred to as the augmented density tensor (ADT). 
In GTEMPO, one first represents each $\gK$ and $\gI_{p}$ as an open boundary GMPS, and then multiplies them together to obtain $\gA$ as an open boundary GMPS (the multiplication is performed on the fly using a zipup algorithm for efficiency~\cite{ChenGuo2024a}). With $\gA$, any multi-time correlation function of the impurity can be calculated straightforwardly following the standard path integral formalism.
However, the strategy used in the existing imaginary-time GTEMPO method is inefficient, as the problem is TTI with anti-periodic boundary condition but open boundary GMPSs are used, and it has been observed that large bond dimensions of the open boundary GMPSs are required, especially for low temperature~\cite{ChenGuo2024b}.

\begin{figure}
  \includegraphics[width=\columnwidth]{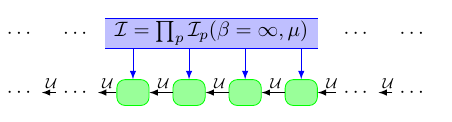} 
  \caption{Schematic illustration of the time-translational invariance of the path integral for the zero-temperature impurity partition function. After integrating out the bath degrees of freedom using the Feynman-Vernon influence functional, only the multi-time impurity state in the temporal domain is left, which is naturally an infinite Grassmann MPS whose unit cell is indicated by the green box.
  Here $\mathcal{U} = e^{-\Himp\delta \tau}$ is the discrete bare impurity propagator with $\delta \tau$ the imaginary-time step size. Here $\beta=\infty$ is the infinite inverse temperature and $\mu$ is the chemical potential.
    }
    \label{fig:tti}
\end{figure}

In fact, the equilibrium QIPs are naturally defined on a torus, where the time axis is in the radial direction and the space axis is in the axial direction (which is infinite since the bath is infinite). This symmetry has been explored in wave-function based MPS methods~\cite{TangWang2020}. In GTEMPO, the spatial direction is integrated out by the Feynman-Vernon IF, and we are left with a one-dimension circle in the time direction only. The time-translational invariance can also be observed from the PI. $\gK$ is time-translationally invariant (TTI) since it only depends on $\Himp$ which is assumed to be time-independent. $\gI_p$ is also TTI since the double integral in its exponent is invariant under any shift of the time indices. The latter property is guaranteed because the hybridization function $\Delta(\tau, \tau')$ in Eq.(\ref{eq:hybrid}) is actually a function of the imaginary-time difference only, e.g., $\Delta(\tau, \tau') = \eta(\tau - \tau')$ for some single-variate function $\eta$. The time-translational invariance of the PI is schematically illustrated in Fig.~\ref{fig:tti}.

Under the framework of the GTEMPO method, 
the TTI property of the IF has been explored in the context of non-equilibrium real-time dynamics to efficiently construct the MPS-IF using essentially a constant number of GMPS multiplications~\cite{GuoChen2024d}. When focusing on the steady state in the infinite-time limit, $\gK$ also becomes TTI and thus the whole ADT can be represented as an infinite GMPS, which has been explored to greatly accelerate the calculation~\cite{GuoChen2024e}. For zero-temperature imaginary-time calculations considered in the work, the PI naturally enjoys a similar symmetry: the whole ADT is TTI and thus can be represented as an infinite GMPS. 
In the following we will show how to construct $\gK$ and $\gI_p$ (thus $\gA$) as infinite GMPSs, and then how to compute the multi-time correlations of the impurity based on $\gA$. 

\subsection{The infinite GTEMPO method}

\begin{figure}
  \includegraphics[width=\columnwidth]{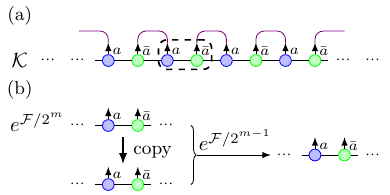} 
  \caption{(a) Constructing the infinite Grassmann MPS representation of the bare impurity dynamics $\gK$, where the dashed rectangle marks the unit cell. (b) Constructing the infinite GMPS representation of the influence functional $\gI$, where $m$ infinite GMPS multiplications is used (only the first step is shown, the remaining steps simply repeat the first step) and the error clearly decreases exponentially fast with $m$ by design. We have used the noninteracting Toulouse model for this schematic demonstration, for which there is only one flavor (therefore the flavor indices are omitted), and the unit cell size is $2$.
    }
    \label{fig:fig1}
\end{figure}
The starting point of the GTEMPO method is to discretize the path integral (here we note a recent development of the tensor network IF method which used a continuous MPS representation of the Feynman-Vernon IF to reduce the time discretization error~\cite{ParkChan2024}). 
$\gI_p$ can be discretized using the quasi-adiabatic propagator path integral (QuaPI) method~\cite{makarov1994-path,makri1995-numerical}, which results in
\begin{align}\label{eq:gI}
\gI_{p} \approx e^{-\sum_{j,k=1}^{N}\abar_{p,j}\Delta_{j,k}a_{p,k}},
\end{align}
where $N = \beta/\delta\tau$ with $\delta\tau$ the imaginary-time step size, $\Delta_{j,k} = \int_{j\delta\tau}^{(j+1)\delta\tau} \dd \tau \int_{k\delta\tau}^{(k+1)\delta\tau} \dd\tau' \Delta(\tau,\tau')$ is the discretized hybridization function.
The bare impurity dynamics $\gK$ can be discretized as
\begin{align}\label{eq:gK}
    \mathcal{K}=&
    \mel*{-\bolda_{,1}}{\Uimp}{\bolda_{,N}} \cdots \mel*{\bolda_{,3}}{\Uimp}{\bolda_{,2}}  \mel*{\bolda_{,2}}{\Uimp}{\bolda_{,1}} , 
\end{align}
where $\bolda_{,k}=\{a_{p,k}, a_{q,k}, \cdots\}$ denotes the set of Grassmann variables (GVs) at time step $k$ for all flavors and $\Uimp = e^{- \Himp\delta \tau}$ is the discrete bare impurity propagator. 
We note the first term on the rhs of Eq.(\ref{eq:gK}) which connects $\bolda_{,1}$ and $\bolda_{,N}$ and encodes the anti-periodic boundary condition. 
In the zero-temperature limit, we have $N = \infty$, thus $\bolda_{,1}$ and $\bolda_{,N}$ will never be actually connected. Moreover, since the the bulk terms $\mel*{\bolda_{,j+1}}{\Uimp}{\bolda_{,j}} $ are all the same, one could represent $\gK$ as an infinite GMPS, such that only the site tensors in a single time step need to be stored and manipulated.
This strategy to build $\gK$ as an infinite GMPS is schematically shown in Fig.~\ref{fig:fig1}(a), which is in parallel with the finite case but the operations only need to be performed on one unit cell.

The strategy to build each $\gI_p$ as an infinite GMPS is more involved, but mostly follows the procedures described in the non-equilibrium setup~\cite{GuoChen2024d}, except that here we use infinite GMPSs instead of open boundary GMPSs. In the following we will sketch the major ideas of these procedures, while the details of each operation can be found in Ref.~\cite{GuoChen2024d}.
We denote the exponent of the discretized IF as $\gF_{p} = -\sum_{j,k}^{\infty}\abar_{p,j}\Delta_{j,k}a_{p,k}$.
First, we expand $\eta_x$ (the discretized version of $\eta(\tau - \tau')$) into the summation of $n$ exponential functions as
\begin{align}\label{eq:prony}
\eta_x \approx \sum_{l=1}^n \alpha_l \lambda_l^{|x|}
\end{align}
using the Prony algorithm~\cite{marple2019digital}, for both $x>0$ and $x<0$. With the optimal values of $\alpha_l $ and $\lambda_l$ determined, we can construct 
$\gF_{p}$ as a GMPS with bond dimension $2n+2$, based on which one can further construct
a first-order GMPS approximation of $e^{\gF_{p}/2^m}$ with bond dimension $2n+1$ using the $\WI$ or $\WII$ methods (or higher-order methods based on them)~\cite{ZaletelPollmann2015}, where $m$ is an integer parameter to control the precision of this first-order approximation (we note that the first-order approximation here is not related to the time discretization error in Eq.(\ref{eq:gI})). Finally, we can obtain the infinite GMPS representation of $\gI_{p} = e^{\gF_{p}}$ using $m$ infinite GMPS multiplications: in the $i$th step one multiplies $e^{\gF_{p}/2^{m-i+1}}$ with itself (we note that this idea is similar to the XTRG algorithm for calculating 1D thermal state~\cite{ChenWeichselbaum2018}). This strategy to build $\gI_{p}$ as an infinite GMPS is schematically illustrated in Fig.~\ref{fig:fig1}(b).

The multiplication of two infinite GMPSs can be done in the same way as the multiplication of two open boundary GMPSs~\cite{ChenGuo2024a}, but only need to be performed for a single unit cell. For two infinite GMPSs with bond dimensions $\chi_1$ and $\chi_2$, the resulting infinite GMPS will have a bond dimension $\chi_1\chi_2$, which needs to be compressed into a new infinite GMPS with some fixed bond dimension $\chi$ (otherwise the computational cost will grow exponentially). 
Importantly, the compression of infinite GMPS can be performed efficiently and stably due to the existence of exact canonical form~\cite{orus2014-practical}.
In this work we use the infinite density matrix renormalization group (IDMRG)~\cite{McCulloch2008} method to iteratively compress the infinite GMPS, following the implementation in the package MPSKit.jl~\cite{MPSKit} (the IDMRG algorithm for infinite GMPS compression is also used in the real-time case~\cite{GuoChen2024e}).
In addition, the hybridization function only contains decaying contributions in the zero-temperature limit as analyzed in Sec.~\ref{sec:PI}, namely $|\lambda_l| < 1$, which indicates that the bond dimension of the infinite GMPS could be much smaller than its finite-temperature counterpart. Thus for zero temperature both problems encountered in previous GTEMPO study of finite-temperature equilibrium AIMs: the growing contribution in the hybridization function and the artificial open boundary condition of the GMPS~\cite{ChenGuo2024b}, can be avoided simultaneously. 
This algorithm is referred to as iGTEMPO as infinite GMPSs are used.

There are several sources of error that could affect the accuracy of iGTEMPO: (1) the time discretization error in Eq.(\ref{eq:gI}), which can be suppressed by using a smaller $\delta\tau$; (2) the MPS bond truncation error during compression, which can be well controlled in infinite MPS algorithms~\cite{OrusVidal2008,StauberHaegeman2018}; (3) the error occurred in the Prony algorithm, characterized by the mean square error
\begin{align}
\pronyerror = \sum_{x} (\eta_x - \sum_{l=1}^n \alpha_l \lambda_l^x)^2,
\end{align}
can be suppressed by using a larger $n$ in principle and it has been shown that one could often find very good approximations with a logarithmically scaling $n$~\cite{croy2009-partial,zheng2009-numerical,hu2010-pade,VilkoviskiyAbanin2023}; (4) the hyperparameter $m$ used to control the precision of the first-order infinite GMPS approximation of $e^{\gF_p/2^m}$, the error of which clearly decreases exponentially fast with $m$ by design. 

\subsection{Calculating multi-time impurity correlations}

\begin{figure}
  \includegraphics[width=\columnwidth]{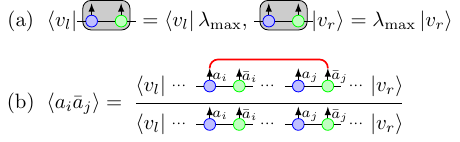} 
  \caption{Algorithm to calculate the Matsubara Green's function based on the infinite GMPS representation of the augmented density tensor at zero temperature, which can be divided into two steps:
  (a) Computing the transfer matrix by integrating the two conjugate Grassmann variables in a unit cell, and then calculating the left and right dominant eigenvectors of it; (b) Identifying a finite window from the infinite ADT depending on the Green's function to be calculated, and then evaluating the Green's function similar to the finite case, but with nontrivial left and right boundary vectors obtained from step (a). We have used the noninteracting Toulouse model for this schematic illustration for briefness.
  The red line means a product operator corresponding to the Grassmann tensor $a_i\abar_j$ which acts on the ADT.
    }
    \label{fig:fig2}
\end{figure}

Once the infinite GMPS representation of $\gA$ has been obtained, one can easily calculate any multi-time correlations of the impurity. For example, the Matsubara Green's function can be calculated as
\begin{align}\label{eq:gf}
-G_j \EqDef& \langle e^{j\delta\tau\Hop} \aop e^{-j\delta\tau\Hop} \adop \rangle = \frac{\int\mea[\boldabar,\bolda] a_j \abar_0 \gA[\boldabar,\bolda]}{\Zimp} .
\end{align}
Based on the infinite GMPS representation of $\gA$, $G_j$ can be computed in two steps (which is the same as the real-time case~\cite{GuoChen2024e}):
(1) Obtaining the transfer matrix by integrating out all the conjugate pairs of GVs in one unit cell, and then calculating the dominant left and right eigenvectors of it, denoted as $\langle v_l\vert $ and $\vert v_r\rangle $ respectively, with dominant eigenvalue denoted as $\lambda_{\max}$; (2) Identifying a finite window from the infinite GMPS representation of $\gA$, and then evaluating the expectation value similar to the finite case, but using $\langle v_l\vert$ and $\vert v_r\rangle$ as left and right boundaries instead of trivial boundaries (Grassmann vacuum $1$). These two steps are schematically shown in Fig.~\ref{fig:fig2}(a,b) respectively. Implementation-wise, this operation boils down to contracting a quasi-2D tensor network of size $(2M+1) \times 4M$ for $M$ orbitals~\cite{ChenGuo2024b} (since we have $2M$ MPS-IFs plus one infinite GMPS for $\gK$, and the unit cell size is $4M$) with open boundary condition, the cost of which roughly scales as $O(\chi^{2M})$.

\section{Numerical results}
To demonstrate the effectiveness and flexibility of the iGTEMPO method, we apply it to study AIMs with increasing complexity, including the noninteracting Toulouse model with a single flavor, the single-orbital AIM with two flavors (spin up and down), and a two-orbital AIM with four flavors. 
We benchmark our iGTEMPO results against exact solutions in the noninteracting case and against the low-temperature CTQMC calculations in the interacting case.
The orderings of the GVs within one unit cell are chosen as $a\abar$, $a_{\uparrow}\abar_{\uparrow} a_{\downarrow}\abar_{\downarrow}$ and $a_{1\uparrow}\abar_{1\uparrow} a_{1\downarrow}\abar_{1\downarrow} a_{2\uparrow}\abar_{2\uparrow} a_{2\downarrow}\abar_{2\downarrow}$ for these three cases respectively, where the time step indices are omitted due to the time-translational invariance.
In all the numerical experiments of this work, we will use a semi-circular bath spectrum density
\begin{align}
J(\varepsilon)=\frac{1}{\pi}D\sqrt{1-(\varepsilon/D)^2},
\end{align}
with $D=1$ ($D$ is used as the unit). The default values of the two hyperparameters, $n$ and $m$, involved in constructing the TTI MPS-IF, are set as $n=25$ (we tend to choose a large enough $n$ such that the error occurred in the Prony algorithm satisfies $\pronyerror < 10^{-4}$) and $m=6$ unless particularly specified (one can also see Ref.~\cite{GuoChen2024d} for the effect of these two hyperparameters on the accuracy of real-time GTEMPO calculations). 

\subsection{Toulouse model}

\begin{figure}
  \includegraphics[width=\columnwidth]{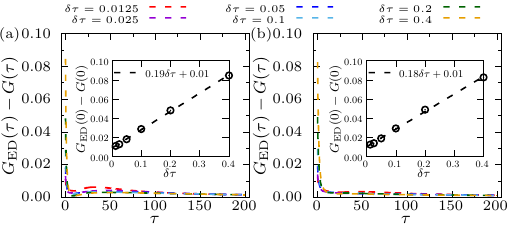} 
  \caption{Error of the zero-temperature Matsubara Green's function $G(\tau)$, compared to the ED results $G_{\rm ED}(\tau)$, as a function of the imaginary time $\tau$ for the Toulouse model for (a) $\varepsilon_d=0$ and (b) $\varepsilon_d=0.5$. The red, purple, blue, cyan, green and yellow dashed lines correspond to iGTEMPO calculations with $\delta\tau=0.0125, 0.025, 0.05, 0.1, 0.2, 0.4$ respectively. The empty circles in both insets show the errors at $\tau=0$ as a function of $\delta\tau$, and the dashed lines are corresponding linear fittings. We have used $n=40$ and $\chi=60$ in all these iGTEMPO calculations.
    }
    \label{fig:fig3}
\end{figure}

We first demonstrate the validity of our method by applying it to study the noninteracting Toulouse model~\cite{mahan2000-many}, for which the impurity Hamiltonian can be written as $\Himp = \varepsilon_d \adop\aop$ with $\varepsilon_d$ the on-site energy.
In the noninteracting limit, one could easily perform exact diagonalization (ED) to obtain numerically exact solutions.
In Fig.~\ref{fig:fig3}(a,b), we plot the error of the zero-temperature equilibrium Matsubara Green's function calculated by iGTEMPO for the Toulouse model, compared to ED results, for $\varepsilon_d=0$ and $\varepsilon_d=0.5$ respectively. For ED we have discretized the bath into equal-distant modes with $\delta\omega/D=0.0002$ and checked that the results have well converged with $\delta\omega$ (the bath discretization error is the only source of error in ED).
For iGTEMPO we have used different values of $\delta\tau$ ranging from $0.4$ to $0.0125$.
We can see that for both values of $\varepsilon_d$, our iGTEMPO results agree increasingly better with ED results for smaller $\delta\tau$. In particular, the errors are mostly concentrated at the first few steps, which is similar to the finite GTEMPO calculations~\cite{ChenGuo2024b} and is likely due to the first-order time discretization error of the IF. However, compared to the finite case, the boundary errors in iGTEMPO seem to be slightly larger.
There is a small jump of error at around $D\tau =30$ when $\delta\tau$ decreases from $0.025$ to $0.0125$ for the half filling case, which may be due to the complex interplay with other error sources that happens for very small $\delta\tau$.
In the insets, we show the errors between the iGTEMPO results and ED results as functions of $\delta\tau$ at $\tau=0$ ($G(0)$ is the average occupation). We can see that the errors are suppressed by decreasing $\delta\tau$. However, the errors against $\delta\tau$ do not extrapolate to zero, which is expected as there exists other sources of errors in the algorithm.

To this end, we discuss an important difference between GTEMPO and iGTEMPO in the choice of $\delta \tau$. In GTEMPO, we generally tend to choose a large $\delta \tau$ as long as the accuracy is not significantly affected, since the computational cost scales linearly with $N = \beta/\delta\tau$. While in iGTEMPO, the computational cost to build the MPS-IFs is essentially independent of $N$, therefore one would like to choose a smaller $\delta\tau $ in this case (meanwhile, a common observation in TEMPO and GTEMPO is that the required bond dimension is smaller for smaller $\delta\tau$~\cite{StrathearnLovett2018,ChenGuo2024b}), the only price to pay is that the cost of computing Green's functions between two times may grow since their distance becomes effectively larger. Despite considerations on the computational cost and the time discretization error, there is another important advantage of using a smaller $\delta\tau$: one could directly calculate the Green's functions on the imaginary-frequency axis instead, which can be expressed as a TTI operator acting on the infinite GMPS representations of the ADT (this calculation can be done for any $\delta\tau$, but for large $\delta\tau$ the error would be huge since we are essentially performing the Fourier transformation of $G(\tau)$). 

\begin{figure}
  \includegraphics[width=\columnwidth]{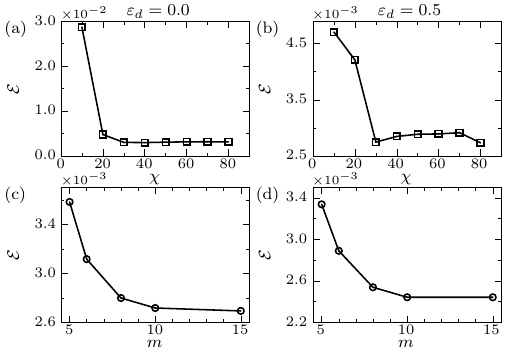} 
  \caption{(a, c) The average error between the zero-temperature Matsubara Green's function calculated by iGTEMPO and that calculated by ED for the Toulouse model with $\varepsilon_d=0$, against (a) the bond dimension $\chi$ and (c) the number of infinite Grassmann MPS multiplications, $m$, used in building the MPS-IF. (b, d) the same plots as in (a, c) but for $\varepsilon_d=0.5$. In (a, b) we have fixed $m=6$, while in (c, d) we have fixed $\chi=60$. We have used $\delta\tau=0.1$ in these simulations.
    }
    \label{fig:fig4}
\end{figure}

In Fig.~\ref{fig:fig4}, we further study the errors in the iGTEMPO calculations against the two hyperparameters $\chi$ and $m$, for $\varepsilon_d=0$ and $\varepsilon_d=0.5$ respectively. We use the \textit{average error}, defined as 
\begin{align}
\mathcal{E}(\vec{x}, \vec{y}) = \sqrt{\frac{||\vec{x} - \vec{y} ||^2}{L}}
\end{align}
to quantify the differences between two vectors $\vec{x}$ and $\vec{y}$ of common length $L$.
Interestingly, from Fig.~\ref{fig:fig4}(a, b) we can see that the average errors between the iGTEMPO results and the ED results decrease and saturate very rapidly against $\chi$: with $\chi=30$ the errors stop decreasing anymore. This is in contrast with the finite GTEMPO calculations under the same bath spectrum density, where it is found that the required bond dimension is already larger than $100$ at $\beta=40$ and keeps growing with the $\beta$~\cite{ChenGuo2024b}. These results also well agree with the intuition that the purely decaying hybridization function at zero temperature would lead to a much lower bond dimension of the underlying infinite GMPS compared to finite temperature. Therefore, although the zero-temperature limit is a hard regime for the CTQMC methods, it could be an easy regime for the iGTEMPO method. The non-monotonic behavior of the average error against $\chi$ in Fig.~\ref{fig:fig4} is likely due to the interplay with other error sources. In Fig.~\ref{fig:fig4}(c, d), we can see that the average errors decrease with larger $m$, which is expected since the first-order discretization error in approximating $e^{-\gF/2^m}$ decreases with larger $m$. Meanwhile, the decrease of error against $m$ is not significant, which means that we can already obtain accurate results with a very small value of $m$.


\subsection{Single-orbital Anderson impurity model}

Now we proceed to study the single-orbital AIM with 
\begin{align}
\Himp = \varepsilon_d \sum_{\sigma} \adop_{\sigma}\aop_{\sigma} +  U\adop_{\uparrow}\adop_{\downarrow} \aop_{\downarrow} \aop_{\uparrow},
\end{align}
where $U$ is the Coulomb interaction and $\sigma \in \{\uparrow, \downarrow\}$ is the electron spin index. We focus on the half-filling scenario with $\varepsilon_d=-U/2$. 

\begin{figure}
  \includegraphics[width=\columnwidth]{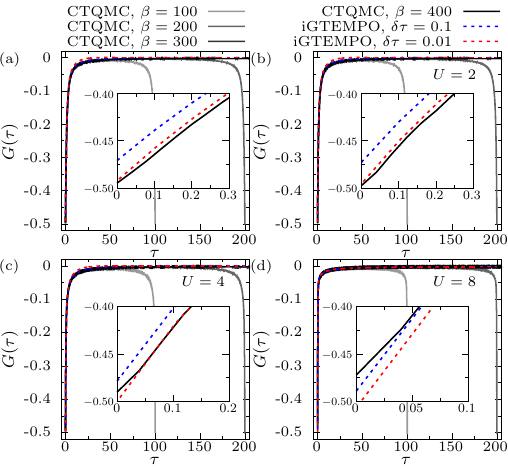} 
  \caption{Zero-temperature Matsubara Green's function $G(\tau)$ as a function of the imaginary time $\tau$ for the single-orbital Anderson impurity model. The four panels show the results for (a) $U/D=1$, (b) $U/D=2$, (c) $U/D=4$ and (d) $U/D=8$ respectively. The blue and red dashed lines are the iGTEMPO results calculated with $\delta\tau=0.1$ and $\delta\tau=0.05$ respectively (we have used $\chi=60$ in both cases). The gray solid lines from lighter to darker are CTQMC results for $\beta=100, 200, 300, 400$ respectively. The insets show the zooms for small $\tau$, where only the iGTEMPO results and the CTQMC results at $\beta=400$ are shown.
    }
    \label{fig:fig5}
\end{figure}

In Fig.~\ref{fig:fig5}, we plot the Matsubara Green's function of the single-orbital AIM calculated by iGTEMPO for $U/D=1,2,4,8$ respectively, with $D\delta\tau=0.1$ (blue dashed lines) and $D\delta\tau=0.01$ (red dashed lines), and $\chi=60$ in both cases. The CTQMC results for $\beta=100, 200, 300, 400$ are also shown as comparisons (for CTQMC calculations we have used $8$ Markov chains and generated $10^7$ samples per chain). We can see that for all different $U$s the iGTEMPO results generally agree well with the CTQMC results: the errors in the bulk are less than $1\%$, for the few initial points the errors are around $5\%$ for $D\delta\tau=0.1$ but become significantly smaller for $D\delta\tau=0.01$ (again there is a small jump of error at around $\tau=20$ for $D\delta\tau=0.01$ and $U/D\leq 6$, similar to Fig.~\ref{fig:fig3}).
In the insets we show the zooms of the initial imaginary-time steps, where we can clearly see that the errors of iGTEMPO are significantly suppressed with a smaller $\delta\tau$.
Interestingly, the iGTEMPO results for $U/D=8$ are already very accurate at the initial imaginary-time steps even with $D\delta\tau=0.1$, which is similar to the observations in the finite-temperature case that the GTEMPO calculations are more accurate for larger $U$~\cite{ChenGuo2024b} (we note that from the inset of Fig.~\ref{fig:fig5}(d) we can see that the CTQMC results for $U/D=8$ have not well converged for small $\tau$ under our given set of parameters, as the half filling condition is clearly violated).

\subsection{Two-orbital Anderson impurity model}

\begin{figure}
  \includegraphics[width=\columnwidth]{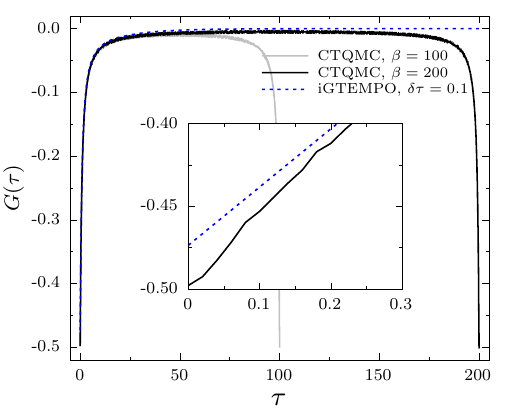} 
  \caption{Zero-temperature Matsubara Green's function $G(\tau)$ as a function of the imaginary time $\tau$ for the two-orbital Anderson impurity model with impurity Hamiltonian in Eq.(\ref{eq:Himp2}). The blue dashed line shows the iGTEMPO results calculated with $D\delta\tau=0.1$ and $\chi=30$. The gray and black solid lines show the CTQMC results calculated for $\beta=100,200$ respectively. The inset shows the zoom for small $\tau$.
    }
    \label{fig:fig6}
\end{figure}

Finally, we study the multiple-orbital AIM with impurity Hamiltonian
\begin{align}\label{eq:Himp2}
    \Himp=&\varepsilon_d\sum_{x, \sigma}\adop_{x, \sigma}\aop_{x, \sigma}
    +U\sum_x\adop_{x,\uparrow}\adop_{x,\downarrow}\aop_{x,\downarrow}\aop_{x,\uparrow} \nonumber \\
    &+(U-2J)\sum_{x\neq y}\adop_{x,\uparrow}\adop_{y,\downarrow}\aop_{y,\downarrow}\aop_{x,\uparrow} \nonumber \\
    &+(U-3J)\sum_{x>y,\sigma}\adop_{x,\sigma}\adop_{y,\sigma}\aop_{y,\sigma}\aop_{x,\sigma} \nonumber \\
    &-J\sum_{x\neq y}(\adop_{x,\uparrow}\adop_{x,\downarrow}\aop_{y,\uparrow}\aop_{y,\downarrow}
    +\adop_{x,\uparrow}\adop_{y,\downarrow}\aop_{y,\uparrow}\aop_{x,\downarrow}),
\end{align}
where the additional parameter $J$ is the Hund’s coupling strength, and we have used $x, y$ to label the orbitals. We focus on the two-orbital case with half filling, under parameters $U=2$, $J=0.5$ and $\varepsilon_d = -(3U-5J)/2$~\cite{Sherman2020}. The previous GTEMPO calculations for this model are limited to the high-temperature regimes, that is, $\beta=10$ for $2$ orbitals and $\beta=2$ for $3$ orbitals, due to the fast growing bond dimensions of the MPS-IFs with $\beta$~\cite{ChenGuo2024b}. 
In Fig.~\ref{fig:fig6}, we show our iGTEMPO calculations using $D\delta\tau=0.1$ and $\chi=30$, with comparisons to the CTQMC results calculated with $\beta=100,200$. We see very good match between the iGTEMPO results and the CTQMC results, except for the initial points where the error is around $5\%$, similar to the single-orbital and noninteracting cases. Importantly, the iGTEMPO results are obtained using a very small bond dimension $30$ only (it can be seen from Fig.~\ref{fig:fig4}(a,b) that the iGTEMPO results have indeed converged at this bond dimension), which is in sharp contrast with previous GTEMPO calculations at finite temperature. Taken into consideration the $O(\chi^{2M})$ scaling of the computational cost of iGTEMPO for $M$-orbital AIMs, it would be promising to extend iGTEMPO to at least the $M=3$ case.

\section{Summary}
In summary, we have proposed an infinite GTEMPO (iGTEMPO) method to solve zero-temperature equilibrium quantum impurity problems. The iGTEMPO method explores the time-translational invariance of the path integral for quantum impurity problems and represents the multi-time impurity state as an infinite Grassmann MPS, which thus benefits from both the GTEMPO method and the infinite MPS algorithm: it is free of the sampling noises and the bath discretization error (while time discretization error still exists), it is free of the sign problem, and its computational cost is independent of $\beta$ as it directly works in the zero-temperature limit. Most importantly, due to the purely decaying hybridization function, the required bond dimension of the infinite GMPS is expected to be much smaller compared to the finite-temperature counterpart, which is confirmed in our numerical simulations. With a small bond dimension $\chi$ and the $O(\chi^{2M})$ scaling of the computational cost of iGTEMPO for $M$-orbital Anderson impurity models, it is promising to use this method as an efficient imaginary-time impurity solver to complement the CTQMC solvers in the zero-temperature regime.


\acknowledgements 
We thank Lei Wang, Shiju Ran for helpful discussions. 
The CTQMC calculations in this work are done using the TRIQS package~\cite{ParcolletSeth2015,SethParcollet2016}.
This work is supported by National Natural Science Foundation of China under Grant No. 12104328. C. G. is supported by the Open Research Fund from State Key Laboratory of High Performance Computing of China (Grant No. 202201-00).

\bibliography{refs}

\end{document}